\documentclass[journal]{IEEEtran}

\usepackage{cite}

\usepackage[pdftex]{graphicx}
\graphicspath{{./}}
\DeclareGraphicsExtensions{.jpg}

\usepackage{amsmath}
\usepackage{mathtools,amssymb}
\usepackage{algorithmic}
\usepackage{array}
\usepackage{url}
\usepackage{arydshln}
\usepackage{tabu}
\usepackage{epsfig}
\usepackage{algorithm}
\usepackage{booktabs}

\usepackage[dvipsnames,svgnames,x11names]{xcolor}
\definecolor{citecolor}{RGB}{119,185,0} 
\usepackage[pagebackref=false,breaklinks=true,letterpaper=true,colorlinks,citecolor=citecolor,bookmarks=false]{hyperref}
\usepackage{soul}
\usepackage[utf8]{inputenc}
\usepackage[small]{caption}
\usepackage{multirow}
\usepackage{comment}
\usepackage{tipa}

\usepackage{pifont}
\def\ie{\emph{i.e.}} 
\def\etal{\emph{et~al.}} 
\def\etc{\emph{etc}}

\newlength\savewidth\newcommand\shline{\noalign{\global\savewidth\arrayrulewidth
  \global\arrayrulewidth 1pt}\hline\noalign{\global\arrayrulewidth\savewidth}}
  
\begin{document}
\title{Subband-based Generative Adversarial Network \\ for Non-parallel Many-to-many Voice Conversion}
\author{Jian Ma, Zhedong Zheng, Hao Fei, Feng Zheng$^*$, Tat-Seng Chua, Yi Yang,~\IEEEmembership{Senior~Member,~IEEE} 
\thanks{$^*$ Corresponding author.}
\thanks{Jian Ma is with Department of Computer Science and Engineering, Southern University of Science and Technology, ShenZhen 518055, China and Faculty of Engineering and Information Technology, University of Technology Sydney, Sydney NSW 2007, Australia. E-mail: jian.ma@student.uts.edu.au
}
\thanks{Feng Zheng is with Department of Computer Science and Engineering, Southern University of Science and Technology, Shenzhen 518055, China. E-mail: f.zheng@ieee.org
}
\thanks{Zhedong Zheng, Hao Fei and Tat-Seng Chua are with School of Computing, National University of Singapore, Singapore 118404. E-mail: zdzheng@nus.edu.sg, haofei37@nus.edu.sg, dcscts@nus.edu.sg}
\thanks{Yi Yang is with School of Computer Science, Zhejiang University, Zhejiang 310058, China. E-mail: yangyics@zju.edu.cn}}

\markboth{Journal of \LaTeX\ Class Files,~Vol.~14, No.~8, August~2015}%
{Shell \MakeLowercase{\textit{et al.}}: Bare Demo of IEEEtran.cls for IEEE Journals}

\maketitle

\begin{abstract}
Voice conversion is to generate a new speech with the source content and a target voice style.
In this paper, we focus on one general setting, \ie, non-parallel many-to-many voice conversion, which is close to the real-world scenario.
As the name implies, non-parallel many-to-many voice conversion does not require the paired source and reference speeches and can be applied to arbitrary voice transfer.
In recent years, Generative Adversarial Networks (GANs) and other techniques such as Conditional Variational Autoencoders (CVAEs) have made considerable progress in this field. 
However, due to the sophistication of voice conversion, the style similarity of the converted speech is still unsatisfactory.
Inspired by the inherent structure of mel-spectrogram, we propose a new voice conversion framework, \ie, Subband-based Generative Adversarial Network for Voice Conversion (SGAN-VC).
SGAN-VC converts each subband content of the source speech separately by explicitly utilizing the spatial characteristics between different subbands.
SGAN-VC contains one style encoder, one content encoder, and one decoder. 
In particular, the style encoder network is designed to learn style codes for different subbands of the target speaker. 
The content encoder network can capture the content information on the source speech. Finally, the decoder generates particular subband content. 
In addition, we propose a pitch-shift module to fine-tune the pitch of the source speaker, making the converted tone more accurate and explainable.
Extensive experiments demonstrate that the proposed approach achieves state-of-the-art performance on VCTK Corpus and AISHELL3 datasets both qualitatively and quantitatively, whether on seen or unseen data.
Furthermore, the content intelligibility of SGAN-VC on unseen data even exceeds that of StarGANv2-VC with ASR network assistance.
\end{abstract}

\begin{IEEEkeywords}
Voice Conversion, Style Transfer, Spectrogram, Generative Adversarial Networks.
\end{IEEEkeywords}

\IEEEpeerreviewmaketitle

\begin{figure}
    \centering
    \includegraphics[width=1\linewidth]{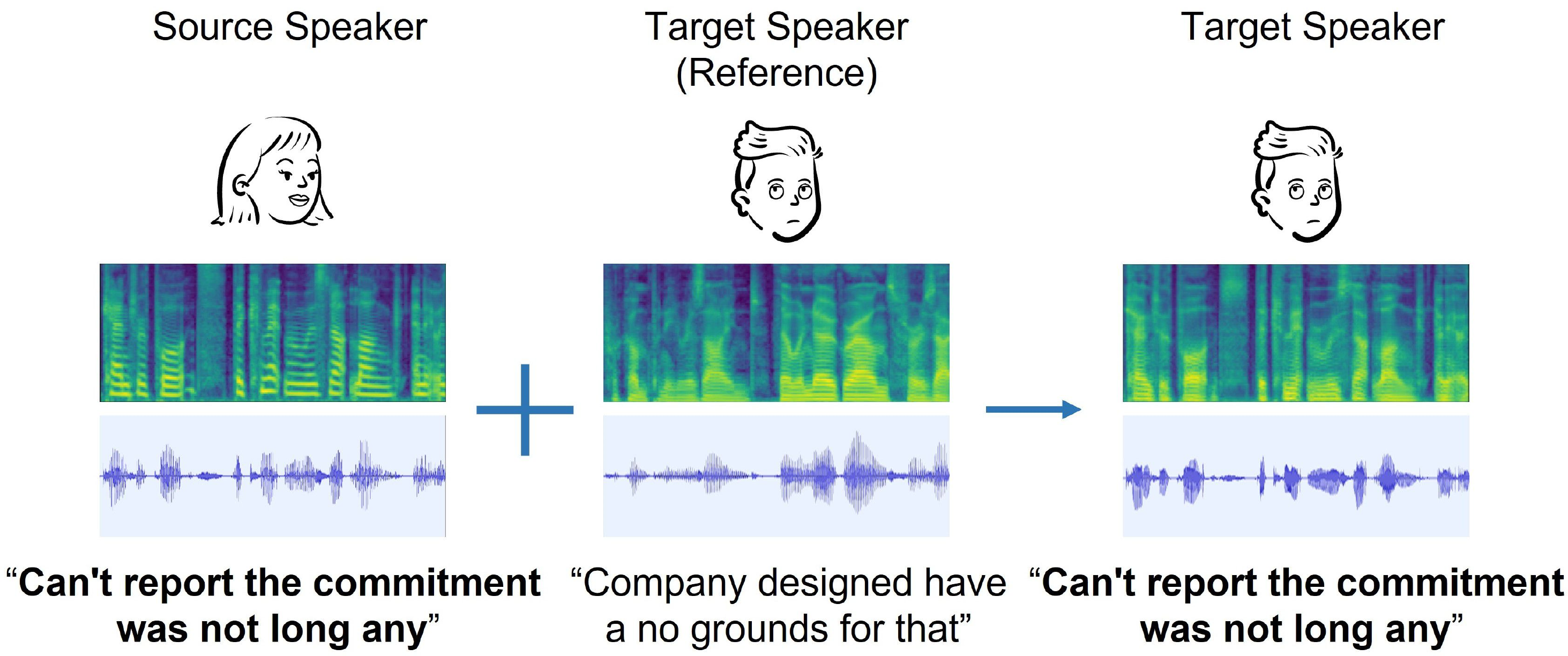}
    \vspace{-.2in}
    \captionof{figure}{ Non-parallel many-to-many voice conversion transfers the speech of the source speaker into the style of an arbitrary target speaker without parallel data, while keeping the source speech content unchanged. Especially when the target speaker does not exist in the training set, it is a challenge to extract style information accurately. We show speakers, mel-spectrograms, speech waveforms, and corresponding text content from top to bottom. The source and target speakers and the converted speech are represented from left to right. The converted speech contains both the style of the target speaker and the text content from the source speaker as if the target speaker utters the sentence.}
    \label{fig:0}
\end{figure}
\maketitle

\section{Introduction} \label{sec: Introduction}

Voice Conversion (VC) aims to generate a new speech with the source content and a reference voice style. 
The generated speech should preserve the source speech content, while transferring to the voice of the target speaker. 
Voice Conversion can be applied to many real-world applications, such as robot voice personalization~\cite{zhao2019voice,savino2021velocity}, voice de-identification~\cite{jin2009voice,pobar2014online}, 
video dubbing \cite{zeng2020talking,wu2021imitating}, speech intelligibility enhancement~\cite{paul20b_interspeech,Wu0RJL21}. 
As depicted in Figure~\ref{fig:0}, the conventional VC process ~\cite{kobayashi2015statistical,kobayashi2014statistical} is defined as that, given two speech fragments from different speakers, the encoder extracts their acoustic features and style information. 
Then the decoder exchanges style information between the source and target speakers to generate a new speech. 
Although existing works have achieved competitive results, due to the collection hardness and annotation costs, the prerequisites are hard to meet in real-world scenarios, including text transcription \cite{sun2016phonetic,DBLP:journals/corr/abs-2010-02434,0001LLJ21} and parallel sentences \cite{tanaka2019atts2s,toda2007voice,fei-etal-2020-cross}. 
Parallel sentences denote that the source speaker and the target speaker speak the same sentences.
Hence, in recent years, researchers start to explore the feasibility of non-parallel many-to-many voice conversion (NPVC)
~\cite{kameoka2018stargan,DBLP:conf/interspeech/KanekoKTH19,qian2019autovc,DBLP:conf/interspeech/LiZM21}.

However, NPVC 
remains challenging in terms of speech naturalness, content intelligibility, and style similarity \cite{DBLP:journals/corr/abs-2008-12527}.  
These challenges mainly come from two aspects.
\textbf{(1) Inherent Voice Discrepancy.} Speech signals can be represented in the frequency domain by the Fourier transform, i.e., a spectrogram. 
In the frequency domain, it is easier to observe the characteristics of the audio signal. 
Generally speaking, in a spectrogram, the vertical axis represents speech frequency and the horizontal axis represents time. 
Affected by the inherent structure of the mel-spectrogram, the converted speech usually converges to the mean timbre of both inputs. 
As a result, the generated speech sounds inconsistent with the style of the target speech.
Moreover, due to differences in physiology, the pitch of female is usually higher than that of male~\cite{2000Clinical}. 
Thus, when conducting voice conversion between different genders, the soprano part of the female may leak into the converted speech,  leading to an unsatisfactory conversion utterance.
\textbf{(2) Extra Voice Losses.} While removing the style information in the source speech, some linguistic content may be discarded, resulting in the incorrect pronunciation of the converted speech. 
Taking the word \textit{win} as an example, the model may convert phoneme [\textipa{I}] to mispronunciation [\textipa{E}]. 
Similarly, [\textipa{n}] may be assigned to incorrect nearby units to become [\textipa{N}], resulting in the wrong pronunciation of [\textipa{wEN}]. [\textipa{wEN}] sounds like the word \textit{Weng}, which is largely different from the original word.
When these mispronunciation and misassignment errors occur multiple times in a speech, it is difficult for humans to accurately understand the content of the source speech.

In the existing works of VC, the entire mel-spectrogram is converted. These holistic conversion methods overlook local detail discrepancies between different mel-spectrograms, where the local details may contain many of the personality characteristics in speech. 
Ignoring the local information will inevitably cause a particular timbre gap with the target speaker.

Inspired by the inherent structure of the mel-spectrogram, we propose a subband-based NPVC framework SGAN-VC.
In particular, a higher pitch is at the upper side of the mel-spectrogram, while a lower pitch is at the bottom side. 
SGAN-VC divides the mel-spectrogram vertically into 4 subbands. 
When generating the converted speech, each subband performs voice conversion independently. 
SGAN-VC simultaneously exchanges local and global information, making the converted speech more similar to the target speaker. 
In addition, to better accommodate the pitch discrepancy between different speakers, we propose a \textbf{pitch-shift module}. 
The pitch-shift module fine-tunes the pitch of the source speaker every time frame. Therefore, SGAN-VC can more accurately convert to the voice of the target speaker.
The converted speech also has a high degree of content intelligibility and audio quality.
Specifically, SGAN-VC is composed of a generator and a discriminator.
The generator contains two encoders, \ie, the content encoder and style encoder, and one decoder. 
The decoder consists of \textbf{Subband-Blocks}.
As the name implies, the content encoder extracts content features in the source speech, while the style encoder learns style embedding of the target mel-spectrogram.
After exchanging stylistic information between different speakers, the decoder generates a converted mel-spectrogram.
Finally, we apply a vocoder to synthesize mel-spectrogram into a sound waveform.

In detail, our style features contain four local parts, where each local feature comes from dividing the global feature into four longitudinally. 
Meanwhile, for the consistency of the overall style, we concatenate the global feature with each local feature. 
Correspondingly, the decoder also has four modules from top to bottom on the vertical axis. 
The four modules of the decoder have the same structure but do not share parameters. Each module generates the content of the corresponding frequency band. 
Finally, we splice the embeddings of the four parts together to obtain the converted spectrogram. The advantage is as follows: 
\textbf{(1)} SGAN-VC can pay attention to global and local information simultaneously. \textbf{(2)} When the source speaker and the target speaker have different vocal ranges, the decoder can decide whether to generate the content of the corresponding subband.
Therefore, the converted speech will not confuse the styles of the source speaker and the reference speaker.
To the best of our knowledge, SGAN-VC is the first framework for subband generation in the field of voice conversion.
Due to differences in gender and age, there are subtle differences in the pitch of different speakers. Even when the same person speaks different sentences, there are discrepancies in pitch and voice intensity.
To mitigate the effects of fine-grained pitch differences, the pitch-shift module vertically shifts the source content features on each frame according to the predicted offset.
Therefore, the generated speech style has a high similarity to the target speaker, while also being highly natural and intelligible.
To better model speech data, SGAN-VC employs two optimization strategies, \ie{}, \textbf{Self-reconstruction Learning} and \textbf{Inter-class Conversion Learning}. Self-reconstruction Learning captures the content and style information of the same speaker to reconstruct itself. Regardless, Inter-class Conversion Learning generates converted speech relying on content and style information provided by different speakers.

We observe that our subband generation strategy is robust and effective in both same-gender and cross-gender voice conversion. 
without the help of text transcription annotations and auxiliary networks, SGAN-VC can achieve state-of-the-art performance.
We conduct experiments on both English and Mandarin datasets. Extensive experiments demonstrate that our method achieves competitive performance. In summary, the main contributions of this paper are as follows:

\begin{itemize}
    \item We propose a simple and effective non-parallel many-to-many voice conversion framework called Subband-based Generative Adversarial Network for Voice Conversion (SGAN-VC). SGAN-VC explicitly utilizes the information of each subband to perform voice conversion respectively. Moreover, it can be trained end-to-end without textual annotations and auxiliary networks, making the model deployment more convenient.
    \item As a minor contribution, we propose a pitch-shift module to predict the frame-level pitch shift, making the conversion of timbre more robust and explainable.
    \item We verify our method on both the English dataset, \ie, VCTK Corpus~\cite{yamagishi2019cstr} and the Mandarin dataset,\ie, AISHELL3~\cite{DBLP:journals/corr/abs-2010-11567}. The proposed method achieves state-of-the-art performance in both source linguistic content preservation and style similarity.
\end{itemize}

The rest of this paper is organized as follows: Section~\ref{sec:related work} briefly summarizes related works on voice conversion. Section~\ref{sec:method} describes the proposed subband-based generative adversarial network in detail. Section~\ref{experiment} comparatively discusses the experimental results, followed by the conclusion in Section~\ref{sec:conclusion}.
Model details are provided in Appendix~\ref{sec:appendices1}.

\section{Related Work}
\label{sec:related work}

\subsection{Generative Adversarial Networks}
Generative Adversarial Networks (GANs) are the representative methods for generation, which are widely employed in many areas, e.g., computer vision~\cite{ChavdarovaF18,Heim19,0019GC0021}, natural language processing~\cite{LiuLYQZL18,YuZWY17,FeiRJ20,fei2021adversarial}, recommendation \cite{CaiHY18,Chen0LJQS19,ZhouXWNKAH21}, \etc.
GANs are pioneered in the field of image generation~\cite{goodfellow2014generative,DBLP:journals/tmm/PengYZL20}, which advances by manipulating the input noise to achieve the desired result~\cite{DBLP:journals/tmm/LiHLWF21,DBLP:journals/tmm/LiDYT21}. 
Abdal~\etal~\cite{abdal2021styleflow} can generate lifelike and detailed faces by editing attributes such as age, gender, and expression. 
In the field of style transfer, Huang~\etal~\cite{huang2017arbitrary} observe that the Adaptive Instance Normalization (AdaIN) structure can well meet the style transfer demand by exchanging the mean and variance of the norm layer between the source and reference samples.
Zheng~\etal~\cite{zheng2019joint} propose DG-Net to transfer the image style between input pairs. Huang~\etal~\cite{huang2021real} apply a similar spirit to face makeup, while Hu~\etal~\cite{hu2020unsupervised} utilize the style feature to remove eyeglasses.
As for the voice generation task, the speech waveform is usually converted from the time domain to the frequency domain by the Fourier Transform.
The spectrogram is then further converted into a mel-spectrogram, which is more suitable for the human auditory system~\cite{DBLP:journals/tmm/ChandrakalaJ20, DBLP:journals/tmm/QianBLOC19}. Multi-Singer \cite{huang2021multi} utilizes the mel-spectrogram when training a singing voice vocoder. Kumar~\etal~\cite{kumar2019melgan} and Yang~\etal~\cite{yang2021multi} also synthesize high-quality sounds through mel-spectrogram. 
For voice style transfer, recent works \cite{9295938,9262021,9023162} find that GANs based on the mel-spectrogram also achieve impressive results.
In this paper, we also deploy GAN as the basic framework.

\subsection{Non-parallel Many-to-many Voice Conversion}
In recent years, deep learning methods have dominated voice conversion.
To obtain high naturalness and intelligibility, previous works utilize some text annotation information or auxiliary networks~\cite{xie2016kl,arik2018neural,0001ZLJ21}, such as Automatic Speech Recognition (ASR), F0 networks, etc. 
Liu~\etal~\cite{DBLP:conf/interspeech/LiuCKHL00M20} employ an ASR network trained with text annotations to recognize phoneme sequences from speech signals.
Based on StarGAN-VC~\cite{kameoka2018stargan}, StarGAN-VC+ASR \cite{DBLP:conf/interspeech/SakamotoTTK21} add the assistance of a pre-trained ASR network to enhance the quality of generated speech.
Le~\etal~\cite{le2021towards} adopt the pre-trained F0 network to promote the similarity with the target speaker. 
Based on the StarGANv2~\cite{choi2020stargan}, StarGANv2-VC~\cite{DBLP:conf/interspeech/LiZM21} supplemented by F0 and ASR network, significantly improves the naturalness and intelligibility of the converted speech. However, due to the limitation of model structure, StarGANv2-VC can only transform the style of the seen data, \ie, all speakers appear in the training set.
Moreover, after removing the auxiliary networks, the performance of StarGANv2-VC drops a lot.
Some works try non-GANs methods.
Lian~\etal~\cite{DBLP:conf/icassp/LianZY22} propose a self-supervised style and content distinguishing model on the Variational Auto-Encoder(VAE) architecture. Akuzawa~\etal~\cite{DBLP:conf/apsipa/AkuzawaOTMM21} utilize a deep hierarchical VAE to achieve high model expressivity and fast conversion speed. Long~\etal~\cite{DBLP:journals/corr/abs-2203-16037} find a suitable location to add the self-attention mechanism to the VAE decoder.
Blow~\cite{serra2019blow} proposes a normalizing flow generation model. Blow first maps the voices of different speakers to the same latent space. Then, the latent expression is converted back to the observation space of the target speaker. AC-VC~\cite{DBLP:conf/asru/RonssinC21} exploits Phonetic Posteriorgrams (PPG) to achieve a high score of 3.5 in terms of naturalness, but sacrifices some speaker similarity. 
To alleviate the reliance of parallel voice conversion on data collection and annotation, researchers begin investigating unsupervised methods. Qian~\etal \cite{DBLP:journals/corr/abs-2106-08519} propose a holistic rhythm transition model without text transcription. 
AUTOVC~\cite{qian2019autovc} and AdaIN-VC~\cite{DBLP:conf/interspeech/ChouL19} conduct zero-shot attempts, \ie, the target speaker is not visible in the training set, denoted as unseen data.
But AUTOVC and AdaIN-VC are flawed in the conservation of the source speech content.

\begin{figure*}[t]
\begin{center}
    \includegraphics[width=1\linewidth]{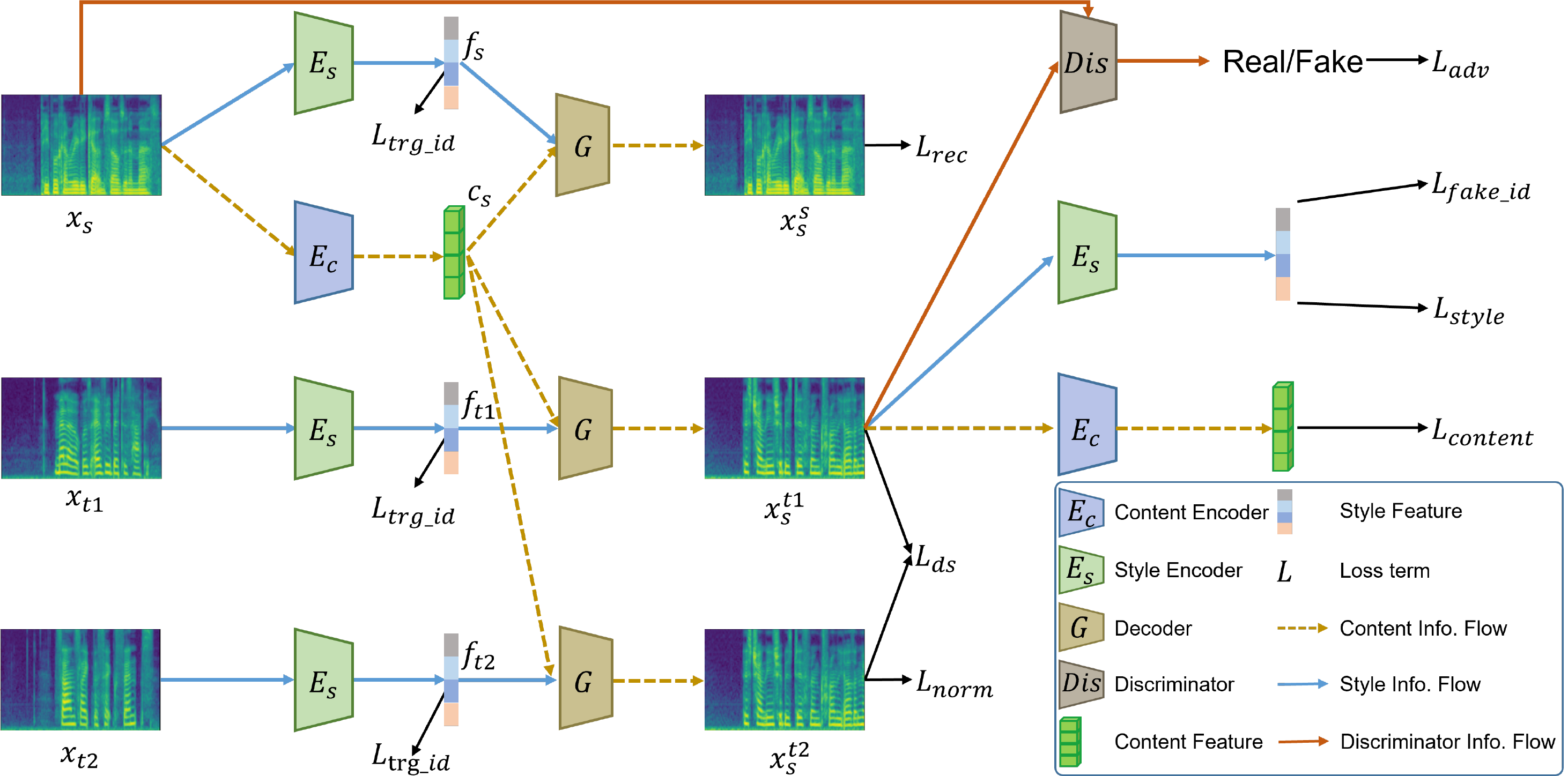}
\end{center}
\vspace{-.1in}
     \caption{A schematic overview of SGAN-VC.
     The \textbf{content encoder} $E_c$ and style encoder $E_s$ are employed to extract content and style features, respectively.
     The generative model $G$ is to generate the converted mel-spectrogram. The discriminator $Dis$ is designed to judge whether the input mel-spectrogram comes from a real sample. Two objectives are enforced in the generative module: (a) Self-reconstruction Learning by the same style and (b) Inter-class Conversion Learning by different target styles.}\label{fig:framework}
\end{figure*}

Unfortunately, previous works fail to pay attention to the differences in the vocal range of different speakers. Converting the entire mel-spectrogram globally is impossible to completely decouple the style and content information. 
Our method exploits the spatial characteristics of different subbands to generate conversion results.
Each subband is converted independently according to the vocal range of different speakers. 
Therefore, the speech generated by SGAN-VC is highly similar to the target speaker.

\section{Methodology} 
\label{sec:method}

As illustrated in Figure~\ref{fig:framework}, SGAN-VC tightly integrates a Generative network and a Discriminative module for voice conversion. The Generator converts the source sample into speech with the target style.
The discriminator judges whether the input mel-spectrogram comes from a real speech sample. The Generator consists of a content encoder, a style encoder and a decoder. To deal with inherent voice discrepancy, we explicitly introduce the Subband-block to aggregate the local feature in the decoder. In particular, our style features are split into four local parts. Each local feature is corresponding to one subband in the speech.  Next, we formally illustrate the problem and symbols.

\subsection{Problem Formulation}
We denote the real mel-spectrogram and class labels as $X=\{x_i\}_{i=1}^N$ and $Y = \{y_i\}_{i=1}^N$, where $N$ is the total number of mel-spectrograms in the dataset. 
In addition, we suppose that the source speaker $S$ and the target speaker $T$ are two variables randomly selected from the speaker pool, respectively. 
Given three real mel-spectrograms $x_s$, $x_{t1}$, and $x_{t2}$, where $x_{t1}$ and $x_{t2}$ are different samples from the same target speaker $T$. 
Our generative module generates a new spectrogram $x^{t1}_{s}$ after exchanging source and target style information.
As shown in Figure~\ref{fig:framework}, the Generator consists of a content encoder $E_c:x_s \rightarrow c_s$, a style encoder $E_s:\{x_{t1} \rightarrow f_{t1}; x_{t2} \rightarrow f_{t2}\}$ and a decoder $G:(x_s, f_{t1}) \rightarrow x^{t1}_{s}$. In particular, we enforce two objectives of the generation module: (1) Self-reconstruction Learning and (2) Inter-class Conversion Learning to make the generated spectrograms controllable.

\noindent\textbf{Self-reconstruction Learning.} As shown in the top two rows of Figure~\ref{fig:framework}, the generative model first learns how to reconstruct itself. It can be expressed as follows:
\begin{equation}
\setlength\abovedisplayskip{2pt}
\setlength\belowdisplayskip{2pt}
x_s^s = G(c_s, f_s) \,.
\end{equation}

\noindent\textbf{Inter-class Conversion Learning.} Unlike self-reconstruction learning, where the spectrogram is reconstructed source content based on its style information. Inter-class conversion focuses on how to generate a spectrogram based on the provided target style.
As shown in the bottom two lines of Figure~\ref{fig:framework}, it can be expressed as follows:

\begin{equation}
x_s^t = G(c_s, f_t).
\end{equation}

\subsection{The Generative Network}
\begin{figure}[t]
\begin{center}
     \includegraphics[width=1\linewidth]{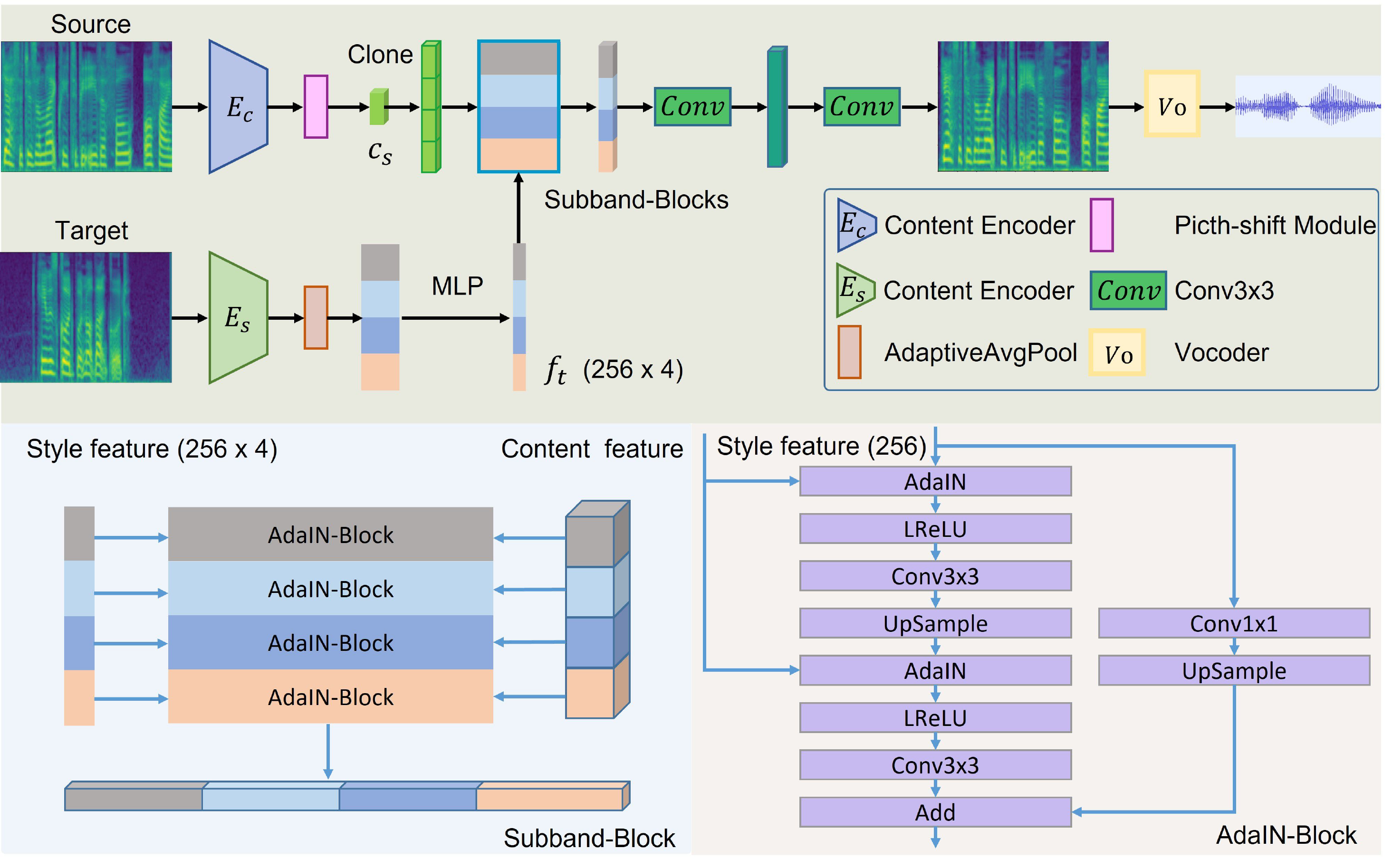}
\end{center}
\vspace{-.1in}
      \caption{The structure of the Subband-Block. A Subband-Block consists of four AdaIN-Blocks. Each part of the style feature from the target speech is fed into the corresponding AdaIN-Block. The Subband-Block exports the respective feature after the style information exchange is completed.
      Finally, the converted spectrogram is synthesized into a speech waveform by a Vocoder.
      }\label{fig:subband-block}
\end{figure}

\noindent\textbf{Content Encoder} is designed to extract the content information of the source speech. 
In particular, Content Encoder is to extract the language-related feature contained in the spectrogram. 
Another function is to remove the speaker-related style information.
As shown in Figure~\ref{fig:subband-block}, we adopt Res-Block as the basic feature extraction module like in StarGANv2~\cite{choi2020stargan}.
Simultaneously, to capture more details, we only downsample twice in the vertical direction and once in the horizontal direction.
A larger feature map enhances the capability of the encoder to perceive fine-grained information.

\vspace{1pt}
\noindent\textbf{Style Encoder} is to remove the content information and obtain the style embedding of the reference speaker.
Our style encoder refers to the ResNet50~\cite{he2016deep}, which has been proven to be a robust image classification model. Since the mel-spectrogram has only one channel, we change the input dimension of ResNet50 to 1.
We focus on the local characteristics of each subband and fine-grained information, so the last downsample of the original ResNet50 is removed. 
Then, as illustrated in Figure~\ref{fig:subband-block}, the feature map is vertically divided into four parts by Adaptive Average Pooling, each part can represent the spatial characteristic of a subband. 
Moreover, to maintain the consistency of the overall style, we collect the global feature from the entire feature map through Average Pooling.
Then we concatenate the global and each local feature to represent the content of each subband, respectively. Finally, an MLP network is used to integrate and reduce the dimensionality of style features.

\vspace{1pt}
\noindent\textbf{Subband-Block.} We apply the content feature and style feature to synthesize the converted mel-spectrogram. 
Therefore, the converted spectrogram contains the language content of the source speech and the style of the target speech. 
To fuse the style feature,  we harness the AdaIN-Block module~\cite{huang2017arbitrary} to exchange the style information.
Motivated by the inherent structure of mel-spectrogram, we design a subband-based module named Subband-Block, which contains four independent AdaIN-Blocks. In particular, as shown in Figure~\ref{fig:subband-block}, we divide the style feature into four subbands. Each AdaIN-Block takes one subband feature as input. In this way, the Subband-Block generates the converted speech from top to bottom according to the frequency band. 
Performing style transfer independently for each frequency band enhances the capacity of the Subband-Block to perceive pitch differences between speakers.
Finally, we convert the fused features of the four subbands into the final mel-spectrogram through two $3\times3$ convolutions.

\vspace{1pt}
\noindent\textbf{Pitch-shift module} is employed to fine-tune the pitch of the source speaker. As shown in Figure~\ref{fig:subband-block}, we employ the pitch-shift module to vertically adjust the content feature to modify the frequency of the source spectrogram.
Pitch-shift module consists of a series of $5\times5$ convolutions and a $1\times1$ convolution.
Finally, the pitch-shift module yields an offset vector with the same time dimension as the content feature. After that, the Tanh activation function is applied to normalize the vector to the $(-1,1)$ interval.
Therefore, the offset vector can represent the displacement of each frame in the mel-spectrogram.
To keep the content information in the source spectrogram, we only perform vertical shifts for each frame.

\subsection{Optimization}
Our goal is to learn the mapping of $x_{s}$ to $x_s^{t1}$ from the source domain to the target domain without parallel data. Referring to DG-Net and StarGANv2-VC, we adopt the following loss function as the objective optimization.

\noindent\textbf{Adversarial loss.} The generator takes a source content feature $c_{s}$ and a style feature $f_{t1}$ and learns how to generate a new spectrogram $x_s^{t1}$. The generator is to cheat the discriminator via the adversarial loss.
\begin{equation}
\begin{aligned}
        L_{adv} &= \mathbb{E}[log(Dis(x_{s}, y_{s}))] \\ &+ \mathbb{E}[log(1-Dis(G(c_{s},f_{t1}), y_{t}))].
\end{aligned}
\end{equation} 
where $Dis(\cdot ; y)$ denotes the output of discriminator for the speaker class $y$ $\epsilon$ $Y$.

\noindent\textbf{ID loss.} For spectrogram category accuracy, we utilize the ID loss for supervision. Meanwhile, to increase the discriminative ability of SGAN-VC, we add an identity classification loss to the target mel-spectrogram.
\begin{align}
   L_{fake\_{id}} & = -\mathbb{E}[log(p(y_{t}|G(c_s,f_{t1})))], \\
    L_{trg\_{id}} & = -\mathbb{E}[log(p(y_s|x_s)) +log(p(y_t|x_{t1})) \nonumber \\
    & + log(p(y_t|x_{t2}))], \\
    L_{id} & = L_{fake\_{id}} + L_{trg\_{id}}.
 \end{align} 
where $p(y_t|x_\cdot)$ denotes the predicted probability of $x_\cdot$ belonging to the class $y_t$.

\noindent\textbf{Style consistency loss.} To ensure that the style of the generated speech is consistent with the target speaker, we apply style consistency loss to guide the generation model.
\begin{align}
    L_{style} = \mathbb{E}[||f_{t1} - E_s(G(c_s, f_{t1}))||_{1}].
\end{align}

\noindent\textbf{Content consistency loss.} Voice conversion changes the style while retaining the linguistic content of the source speech.
The content consistency loss is to ensure that the content of the generated speech is the same as that of the source speech.
\begin{align}
    L_{content} = \mathbb{E}[||c_s - E_c(G(c_s, f_{t1}))||_{1}].
\end{align}

\noindent\textbf{Style diversification loss.} Minimizing the style diversity loss encourages different samples with discriminative styles.
\begin{align}
    L_{ds} = - \mathbb{E}[||G(c_s, f_{t1}) - G(c_s, f_{t2})||_{1}].
\end{align}

\noindent\textbf{Norm consistency loss.} 
The absolute column-sum norm for a mel-spectrogram can represent the degree of the sound energy, and then judge the state of speech/silence. Like StarGANv2-VC~\cite{DBLP:conf/interspeech/LiZM21}, we utilize norm consistency loss $L_{norm}$ to retain the speech/silence status of the source speech. Define $m$ is the index of the $m^{th}$ frame in $x$. The norm consistency loss is given by:
\begin{align}
    L_{norm} = \mathbb{E}[||||x_s[m]||_1 - ||G(c_s,f_{t2})[m]||_1||_1].
\end{align}
where $||x[m]||_1$ and $||G(c,f_{t2})[m]||_1$ represent the absolute column-sum norm of the $m^{th}$ frame of source and converted mel-spectrograms, respectively.

\noindent\textbf{Reconstruction loss.} Based on the spirit of auto-encoder, we motivate  that source speech can be reconstructed based on its own content and style feature.
\begin{align}
    L_{rec} &= \mathbb{E}[||x_s - G(c_s,f_s)||_{1}].
\end{align}

\noindent\textbf{Full generator objective.} Our full generator objective function can be summarized as follows:
\begin{equation}
\begin{aligned}
    L_{total}(E_{c}, E_{s}, G) = \lambda_{adv}L_{adv} + \lambda_{id}L_{id} + \lambda_{style}L_{style} \\
 +  \lambda_{content}L_{content} + \lambda_{ds}L_{ds}  + \lambda_{norm}L_{norm} + \lambda_{rec}L_{rec}.
\end{aligned}
\end{equation} 
where $\lambda_{adv}$, $\lambda_{id}$, $\lambda_{style}$, $\lambda_{content}$, $\lambda_{ds}$, $\lambda_{norm}$ and $\lambda_{rec}$ are hyperparameters for each term. Besides, the discriminator is update by $-\lambda_{adv}L_{adv}$.

\setlength{\tabcolsep}{8pt}
\begin{table}
\small
\begin{center}
\begin{tabular}{lccccc}
\shline
Method & Seen & ASR & Quality$\uparrow$ & Similarity$\uparrow$ \\
\hline
 Ground truth & - & - & 4.800 & - \\
\cdashline{1-5}
 StarGANv2-VC & $\checkmark$ & $\times$ & 1.619 & 2.128 \\
 StarGANv2-VC & $\checkmark$ & $\checkmark$ & 2.503 & 2.745 \\
 SGAN-VC & $\times$ & $\times$ & 2.849 & 1.963 \\
 SGAN-VC & $\checkmark$ & $\times$ & \textbf{3.248} & \textbf{3.173} \\
\shline
\end{tabular}
\end{center}
\vspace{-.1in}
\caption{Qualitative evaluation on the VCTK Corpus test set. }
\label{table:Subjective}
\end{table}

\setlength{\tabcolsep}{5pt}
\begin{table*}
\small
\begin{center}
\begin{tabular}{lcc|cccc|cccc}
\shline
\multirow{2}{*}{Methods} & \multirow{2}{*}{Seen} & \multirow{2}{*}{ASR} & \multicolumn{4}{c}{VCTK} & \multicolumn{4}{c}{AISHELL3-84} \\
\cmidrule(r){4-11}
  & & & pMOS$\uparrow$ &  CLS$\uparrow$ &  CER$\downarrow$ & $m{F0_{diff}}\downarrow$  & pMOS$\uparrow$ &  CLS$\uparrow$ &  CER$\downarrow$ & $m{F0_{diff}}\downarrow$ \\
\hline
 Ground truth & - & - & 3.484 & 96.60 \% & 5.27 \% & - &  3.122 & 99.79 \% & 2.52 \% & - \\
\cdashline{1-11}
AUTOVC & $\times$ & $\times$ & 3.031 & 2.21 \% & 74.18 \% & 16.87 & 3.191 & 1.78 \% & 87.15 \%  & 14.45 \\
AdaIN-VC & $\times$ & $\times$ & 3.573 & 76.48 \% & 60.58 \% & 3.27 & 3.138 & 89.68 \% & 64.59 \%  & 5.46 \\
SGAN-VC-no-Pitch & $\times$ & $\times$ & 3.531 & 26.10 \% & \textbf{17.74 \%} & 4.18 & 3.141 & 39.58 \% & 26.93 \%  & 6.41\\
 SGAN-VC & $\times$ & $\times$ & 3.595 & 27.70 \% & 25.42 \% & 5.13 & 3.130 & 57.08 \% & 25.75 \% & 6.66 \\
 \cdashline{1-11}
 AUTOVC & $\checkmark$ & $\times$ & 3.027 & 86.49 \% & 73.36 \% & 17.22 & 3.055 & 94.25 \% & 101.03 \% & 10.62 \\
 AdaIN-VC & $\checkmark$ & $\times$ & 3.616
 & 73.05 \% & 68.95 \% & 2.90 & 3.090 & 99.06 \% & 91.19 \% & 4.96 \\
 StarGANv2-VC & $\checkmark$ & $\times$ & 3.665 & 94.10 \% & 35.22 \% & 12.12 & 3.155 & 92.92 \% & 69.05 \% & 5.52 \\
 StarGANv2-VC & $\checkmark$ & $\checkmark$ & \textbf{3.792} & 94.80 \% & 18.09 \% &  9.78 & 3.073 & 82.40 \% & \textbf{24.09} \% & 6.52 \\
 SGAN-VC-no-Pitch & $\checkmark$ & $\times$ & 3.468 & 95.00 \% & 24.28 \% & 2.81 & 3.188 & 99.27 \% & 35.95 \% & 4.07 \\
 SGAN-VC & $\checkmark$ & $\times$ & 3.479 & \textbf{97.60 \%} & 20.78 \% & \textbf{1.88} & \textbf{3.206} & \textbf{99.90 \%} & 35.61 \% & \textbf{3.67} \\
\hline

\end{tabular}
\end{center}
\vspace{-.1in}
\caption{Quantitative results on VCTK Corpus and AISHELL3-84 test set. 
"Seen" in the first row indicates whether the speakers are present in the training set. 
"ASR" demonstrates whether StarGANv2-VC has added the assistance of the ASR network. 
All metrics are evaluated on the random source and target pairs. 
The test samples of VCTK Corpus and AISHELL3-84 contain 1000 and 960 utterances, respectively. "SGAN-VC-no-Pitch" denotes SGAN-VC without the pitch-shift module.}
\label{table:object vctk}
\end{table*}

\section{EXPERIMENT} \label{experiment}

\subsection{Datasets}
We mainly evaluate SGAN-VC on two datasets: VCTK Corpus~\cite{yamagishi2019cstr} and AISHELL3~\cite{DBLP:journals/corr/abs-2010-11567}, also including evaluation of seen and unseen data.
\textbf{Seen data} implies that the training set contains speakers in the test set. On the contrary, \textbf{unseen data} indicates that the speakers in the test set do not appear in the training set.
We strongly encourage readers to listen to the audio samples.\footnote{\url{https://hechang25.github.io/SGAN-VC}}

\noindent\textbf{VCTK Corpus} \cite{yamagishi2019cstr} contains approximately 44 hours of speech recordings from 109 speakers.
VCTK Corpus contains 47 male speakers and 62 female speakers, with a relatively balanced gender ratio. 
For a fair comparison, we first utilize the same 20 speakers reported in \cite{DBLP:conf/interspeech/ChouYLL18, DBLP:conf/interspeech/LiZM21} for the seen data experiment, called VCTK20.
The discrepancy from \cite{DBLP:conf/interspeech/ChouYLL18, DBLP:conf/interspeech/LiZM21} is that all our audio fragments are randomly sampled from the original VCTK Corpus.
Therefore, the pMOS of ground truth is lower than that reported in the StarGANv2-VC~\cite{DBLP:conf/interspeech/LiZM21}.
For data balance, each speaker has the same number of audio samples.
Regarding the test set, we select $5$ males and $5$ females from VCTK20. 
Each speaker contains $50$ samples that do not exist in the training data.
Ultimately, in the training set, for the seen data, \ie{}, VCTK20, each speaker contains $150$ samples. 
For the unseen data experiment, our training set applies all the speakers of the original VCTK Corpus except the $10$ speakers in the test set.
To reduce training time, each speaker retains only $50$ samples.

\vspace{2pt}
\noindent\textbf{AISHELL3}~\cite{DBLP:journals/corr/abs-2010-11567} is a large-scale and high-fidelity multi-speaker Mandarin speech corpus. 
AISHELL3 contains roughly $85$ hours of recordings produced by 218 Chinese speakers (consisting of 176 females and 42 males). A total of $88,035$ utterances are recorded.
Due to the unbalanced gender ratio of AISHELL3, we employ all male speakers and a randomly selected array of 42 female speakers as our evaluation dataset, named \textbf{AISHELL3-84}.
Likewise, $5$ male and $5$ female speakers are randomly selected in AISHELL3-84 as the final test set. 
Due to the short duration of many audio clips in AISHELL3-84, we remove samples with durations less than 2.5s.
Eventually, each speaker possesses $50$ audio clips in the training set. 
In the test set, each speaker  contains $48$ samples.

\subsection{Training Details}
For data processing, we first resample all audio clips to $22.050$ kHz and convert the original speech waveform to a mel-spectrogram. 
The FFT size and hop size are $1024$ and $256$ respectively.
The scale of the mel-spectrogram is 80-bin.
Then we crop the width of the mel-spectrogram to $224$, and fill it with $0$ if it is not enough.
Eventually, the shape of each mel-spectrogram is $(1, 80, 224)$.

\setlength{\tabcolsep}{7pt}
\begin{table}[!t]
\small
\begin{center}
\begin{tabular}{lccc}
\shline
Method & Type & CLS$\uparrow$ &  CER$\downarrow$ \\
\hline
 \multirow{2}{*}{Ground truth} & F & 93.60 \% & \multicolumn{1}{r}{5.69\%}   \\
         & M & 99.60 \% & \multicolumn{1}{r}{4.85\%} \\
\shline         
\multirow{4}{*}{StarGANv2-VC-noASR} & F2F & 95.36 \% & 30.69 \% \\
                     & F2M & 94.44 \% & 39.85 \% \\ 
                     & M2F & 92.74 \% & 34.89 \% \\ 
                     & M2M & 93.65 \% & 36.25 \% \\
\hline
\multirow{4}{*}{StarGANv2-VC-ASR} & F2F & 94.29 \% & \textbf{17.11} \% \\
                         & F2M & 97.01 \% & \textbf{19.32} \% \\ 
                         & M2F & 90.60 \% & \textbf{18.38} \% \\ 
                         & M2M & 97.22 \% & \textbf{17.78} \% \\
\hline
\multirow{4}{*}{SGAN-VC-Unseen} & F2F & 30.95 \% & 22.75 \% \\
                         & F2M & 24.50 \% & 30.92 \% \\ 
                         & M2F & 28.11 \% & 26.34 \% \\ 
                         & M2M & 27.20 \% & 21.70 \% \\
\hline
\multirow{4}{*}{SGAN-VC-Seen} & F2F & \textbf{95.64} \% & 19.60 \% \\
                         & F2M & \textbf{99.20} \% & 22.85 \% \\ 
                         & M2F & \textbf{95.58} \% & 21.43 \% \\ 
                         & M2M & \textbf{100.00} \% & 19.29 \% \\
\hline
\end{tabular}
\end{center}
\vspace{-.1in}
\caption{Results of four conversion types for different methods on the VCTK Corpus test set. $M$ and $F$ represent male and female, respectively. $F2M$ means that the source and target speakers are female and male, respectively, and others are similar.}
\label{table:gender_cls}
\end{table}

In the training phase, we first convert the sound signal into a mel-spectrogram. Then the source and target spectrograms are fed into the content encoder and style encoder respectively.
Finally, the decoder exports a mel-spectrogram with the source content and the target style.
The discriminator differentiates whether the generated spectrogram is close to the real data.
In the inference stage, the Vocoder converts the mel-spectrogram generated by the Generator into a sound waveform.
We train our model for $100$ epochs with a batch size of $16$, about $2.6$ second long audio segments. We employ AdamW~\cite{DBLP:conf/iclr/LoshchilovH19} optimizer with a learning rate of $0.0001$.
For data augmentation, we mainly use Time Warping and Frequency Masking proposed in \cite{DBLP:conf/interspeech/ParkCZCZCL19}. 
The style encoder is first pre-trained on the same training set.
Drawing on DG-NET and StarGANv2-VC, we set $\lambda_{adv}=2$, $\lambda_{id}=0.5$, $\lambda_{style}=5$, $\lambda_{content}=10$, $\lambda_{norm}=1$, $\lambda_{ds}=1$, $\lambda_{rec}=5$.

\setlength{\tabcolsep}{7pt}
\begin{table}
\small
\begin{center}
\begin{tabular}{lcccc}
\shline
Num of parts & pMOS$\uparrow$ &  CLS$\uparrow$ &  CER$\downarrow$ & $mF0_{diff}\downarrow$ \\
\shline
 Ground truth & 3.484 & 96.6 \% & 5.27 \% & -  \\
\cdashline{1-5}
 num=1 & 3.432 & 97.2 \% & 22.70 \% & 3.31  \\
 num=2 & 3.453 & 96.7 \% & 39.33 \% & 3.91  \\
 num=3 & 3.462 & 97.1 \% & \textbf{18.62} \% & 3.57  \\
 num=4 & \textbf{3.479} & \textbf{97.6} \% & 20.78 \% & \textbf{1.88} \\
 num=5 & 3.464 & 96.1 \% & 21.30 \% & 2.69\\
\hline

\end{tabular}
\end{center}
\vspace{-.1in}
\caption{The effect of the number of parts on the VCTK20 dataset. "num" indicates how many subbands we divide the mel-spectrogram.}
\label{table:num of part effects}
\end{table}

\subsection{Evaluations}
We evaluate our model with both qualitative and quantitative metrics. The ablation study is mainly conducted  on the VCTK20 dataset. Due to the time costs and expenses, we only conduct a qualitative experiment on the VCTK test set.

\setlength{\tabcolsep}{12pt}
\begin{table*}
\small
\begin{center}
\begin{tabular}{lccccc|cc}
\shline
\multirow{2}{*}{ID Number} & \multirow{2}{*}{Gender} & \multicolumn{2}{c}{StarGANv2-VC} & \multicolumn{4}{c}{SGAN-VC} \\
\cmidrule(r){3-4}\cmidrule(r){5-8}
& & noASR & ASR & Seen-no-Pitch & Seen-Pitch & Unseen-no-Pitch & Unseen-Pitch \\

\shline
 1 & F & 13.15 & 21.06 & 2.92 & 2.97 & 2.47 & 3.99 \\
 2 & F & 16.30 & 14.46 & 1.61 & 0.85 & 6.93 & 7.91 \\
 3 & F & 16.49 & 12.90 & 6.45 & 0.88 & 3.35 & 4.37 \\
 4 & F & 11.34 & 14.96 & 1.02 & 0.44 & 4.42 & 1.50 \\
 5 & F & 15.03 & 12.45 & 5.42 & 2.66 & 6.80 & 9.75 \\
 \cdashline{1-8}
 6 & M & 18.64 & 7.73 & 1.43 & 3.79 & 0.21 & 4.51 \\
 7 & M & 17.90 & 2.11 & 1.33 & 0.90 & 4.62 & 2.20 \\
 8 & M & 1.05 & 1.56 & 6.01 & 5.39 & 6.07 & 6.98 \\
 9 & M & 5.11 & 4.50 & 1.38 & 0.96 & 0.57 & 3.19 \\
 10 & M & 6.25 & 6.09 & 0.57 & 0.00 & 6.33 & 6.91 \\
\cdashline{1-8}
 \textbf{m$F0_{diff}$} & - & 12.12 & 9.78 & 2.81 & \textbf{1.88} & \textbf{4.18} & 5.13 \\
\hline

\end{tabular}
\end{center}
\vspace{-.1in}
\caption{Comparison of $F0_{diff}$ of different methods on the VCTK Corpus test set. 
We replace the ID information of the speakers with numbers $1-10$. All values in the table are in Hz.}
\label{table:vctk pitch difference}
\end{table*}

\noindent\textbf{Qualitative metric.}
As mentioned above, we randomly select $5$ male and $5$ female speakers as our target speakers in the three datasets VCTK20, VCTK, and AISHELL3-84. To ease the comparison, VCTK20 and VCTK Corpus have the same test data.
The source and target speakers are randomly selected from all $10$ speakers.
Since the previous works require subjects to score audio clips, the audio samples are given $1-5$ points in each evaluation metric. 
The mean opinion score (MOS) of audio samples serves as the final comparison basis.
However, the scoring rule of MOS is complicated and is easily influenced by historical data.
Therefore, we adopt a simple and effective qualitative evaluation method.
We make a questionnaire, where each conversion set contains source speech, target speech, state-of-the-art model, and our conversion results, as well as the text information of source speech. 
We require volunteers to rank them in terms of \textbf{quality} of speech and style \textbf{similarity} to the target speech.
The \textbf{quality} is mainly scored from three aspects: noise level, content intelligibility, and speech naturalness.
There are four conversion samples for each set of conversion pairs, \ie, StarGANv2-VC with ASR network, StarGANv2-VC without ASR network, and SGAN-VC network trained on unseen and seen data, respectively.
For the evaluation of quality, due to the participation of source speech, the best score is $5$ and the worst is $1$. 
For the evaluation of \textbf{similarity}, the score range is $(1,4)$.
The higher of both of these two metric, the better the results. 
Finally, we average the scores of all converted samples.

\setlength{\tabcolsep}{12pt}
\begin{table*}
\small
\begin{center}
\begin{tabular}{lccccc|cc}
\shline
\multirow{2}{*}{ID Number} & \multirow{2}{*}{Gender} & \multicolumn{2}{c}{StarGANv2-VC} & \multicolumn{4}{c}{SGAN-VC} \\
\cmidrule(r){3-4}\cmidrule(r){5-8} & & noASR & ASR & Seen-no-Pitch & Seen-Pitch & Unseen-no-Pitch & Unseen-Pitch \\
\shline
 1 & F & 10.72 & 10.05 & 4.98 & 4.53 & 2.10 & 4.22 \\
 2 & F & 2.88 & 4.17 & 8.96 & 9.71 & 6.79 & 10.85 \\
 3 & F & 2.02 & 9.04 & 4.72 & 3.15 & 5.09 & 4.19 \\
 4 & F & 18.17 & 15.23 & 3.74 & 2.35 & 12.00 & 2.67 \\
 5 & F & 1.00 & 0.55 & 1.52 & 5.09 & 3.73 & 5.96 \\
 \cdashline{1-8}
 6 & M & 2.97 & 3.67 & 2.67 & 2.41 & 2.33 & 7.66 \\
 7 & M & 2.97 & 4.45 & 0.04 & 0.33 & 11.91 & 6.97 \\
 8 & M & 2.48 & 8.87 & 8.80 & 2.70 & 6.04 & 5.98 \\
 9 & M & 3.67 & 9.00 & 0.76 & 4.21 & 4.13 & 11.29 \\
 10 & M & 8.33 & 0.16 & 4.53 & 2.29 & 10.00 & 6.79 \\ 
 \cdashline{1-8}
 $mF0_{diff}$ & - & 5.52 & 6.52 & 4.07 & \textbf{3.68} & \textbf{6.41} & 6.66 
\\
\hline

\end{tabular}
\end{center}
\vspace{-.1in}
\caption{Comparison of $F0_{diff}$ of different methods on the AISHELL3-84 dataset. We replace the ID information of the speaker with numbers 1-10. All values in the table are in Hz.}
\label{table:AISHELL3-84 pitch difference}
\end{table*}

\noindent\textbf{Quantitative metric.}
We employ the predicted mean opinion score (\textbf{pMOS}) from MOSNet~\cite{DBLP:conf/interspeech/LoFHWYTW19}, \textbf{$mF0_{diff}$}, classification accuracy (\textbf{CLS}), and automatic speech recognition (ASR) accuracy for quantitative evaluation.
Similar to~\cite{DBLP:conf/interspeech/PolyakWAT20} and StarGANv2-VC~\cite{DBLP:conf/interspeech/LiZM21}, we adopt ResNet as the classifier. 
Besides, we train ResNet on all speaker data for the VCTK Corpus and AISHELL3-84 Dataset, since the style characteristics of some speakers are similar. 
Training only on the selected 10 test speakers will result in some inaccurate conversion results being misclassified, leading to falsely high accuracy. 
For intelligibility evaluation, we adopt the high-performance ASR open-source toolkit WeNet \cite{DBLP:conf/interspeech/YaoWWZYYPCXL21}. 
For the VCTK Corpus dataset, we utilize a pre-trained model on LibriSpeech~\cite{panayotov2015librispeech} dataset.
For the AISHELL3-84 dataset, we employ the pre-trained model on multiple fusion data sets such as AISHELL-1~\cite{bu2017aishell}, AISHELL-2~\cite{DBLP:journals/corr/abs-1808-10583}, THCHS-30~\cite{DBLP:journals/corr/WangZ15e}, and Primewords Chinese Corpus~\cite{primewords_201801}, etc.
Character Error Rate (CER) is operated as an intelligibility evaluation metric for the AISHELL3-84 and VCTK Corpus datasets.
Moreover, when the difference between the two conversion results of the same target is small, it is difficult for the human ear to distinguish the discrepancy.
Therefore, we propose to exploit the average F0 difference between the conversion sample and the target speaker as an objective evaluation, which can more effectively evaluate the accuracy of the style similarity:
\begin{align}
    F0_{diff} &= ||F0_{x_s^t} - F0_{x^t}||_1.\\
    mF0_{diff} &= \mathbb{E}[F0_{diff}]
\end{align}

\subsection{Comparison with the State-of-the-arts}

\noindent\textbf{Qualitative Results.} Table~\ref{table:Subjective} represents the ranking of the converted samples perceived by the human ear.
We select $5$ male and $5$ female speakers and voice-converted each with everyone else, resulting in a total of $100$ converted samples. 
We invite $18$ volunteers to judge the speech quality and similarity. 
The averages of these metrics are reported in the table. 
The score represents the high or low ranking, \ie{}, the higher the score, the higher the ranking.
For \textbf{seen data}, SGAN-VC outperforms StarGANv2-VC in both quality and similarity metrics.
Since the volume of data in the unseen experiment far exceeds that of the seen, the speech quality of our method also surpasses StarGANv2-VC with an ASR network on unseen data.
In terms of similarity, it is worth mentioning that the SGAN-VC based on \textbf{unseen data} is close to the StarGANv2-VC without the ASR network.
It shows that SGAN-VC still has a strong generalization capability for unseen speakers.

\vspace{3pt}
\noindent\textbf{Quantitative Results.}
In Table~\textbf{\ref{table:object vctk}}, AdaIN-VC achieves the highest results on the CLS metric of the unseen experiment. However, the goal of voice conversion is to generate utterances with source speech content and target voice. The CER of AdaIN-VC is as high as $64.59\%$, which causes severe damage to the intelligibility of the source speech. Similarly, AUTOVC cannot meet the requirements of speech intelligibility.
Therefore, this paper mainly compares and analyzes the two methods of SGAN-VC and StarGANv2-VC.
As shown in Table~\textbf{\ref{table:object vctk}}, in terms of \textbf{pMOS}, both StarGANv2-VC and SGAN-VC can generate high-quality speech close to natural speech.
Specifically, SGAN-VC is slightly lower than StarGANv2-VC on VCTK Corpus and slightly higher than StargANv2-VC on the AISHELL3-84 dataset.
As can be seen from ~\textbf{\ref{table:object vctk}}, the pMOS of VCTK Corpus is much higher than that of AISHELL3-84.
The pMOS of StarGANv2-VC drops a lot on different datasets. However, the performance of SGAN-VC is relatively stable, even exceeding the ground truth by about 0.08.
SGAN-VC reduces the requirements for audio quality.
Compared with high-fidelity speech, we are more concerned with the similarity of timbre and the intelligibility of content.
As for \noindent\textbf{CLS}, in experiments on seen data, we achieve $+2.80\%$ and $+6.98\%$ improvement over StarGANv2-VC on VCTK Corpus and AISHELL3-84, respectively, suggesting that the samples we generate have a higher timbre similarity to the target speaker.
The assistance with the ASR network can be considered to utilize text annotation information to some extent. 
As shown in Table~\textbf{\ref{table:object vctk}}, when it comes to the \textbf{CER}, after removing the ASR network, the CER of StarGANv2-VC increases from $18.09\%$ to $35.22\%$. 
Even though our method does not employ any text transcription, the SGAN-VC trained on the seen data is only $2.69\%$ inferior to the StarGANv2-VC with the ASR network on the VCTK Corpus. However, when on the unseen data, due to the greatly increased training data, SGAN-VC surpasses StarGANv2-VC by $0.35\%$.
Similarly, on AISHELL3-84, without the assistance of the ASR network, the CER of StarGANv2-VC raises dramatically by $44.96\%$.
After increasing the amount of data, the CER of SGAN-VC drops by about 10 percentage points.
We can witness that SGAN-VC has the capacity to retain the source language content.

For $mF0_{diff}$, SGAN-VC is significantly smaller than other models, and even the $mF0_{diff}$ of several speakers are less than 1Hz, which further illustrates the effectiveness of SGAN-VC. Meanwhile, even in the unseen data, our $mF0_{diff}$ is smaller than the seen result of StarGANv2-VC, which explains why our classification accuracy is much lower than StarGANv2-VC, but close to StarGANv2-VC in qualitative test results (Table~\textbf{\ref{table:Subjective}}).

\subsection{Comparison of different conversion types}
\noindent\textbf{Table~\ref{table:gender_cls}} displays the results for different transformation types in the test set of VCTK Corpus.
Because of the small number of samples per conversion type, $mF0_{diff}$ is not calculated here.
After obtaining high-quality conversions, we focus more on style similarity and content intelligibility.
Therefore, we mainly calculate the \textbf{CLS} and \textbf{CER}.
$F2F$ and $M2M$ stand for same-gender conversion. $F2F$ is a conversion between female speakers, and $M2M$ is the conversion between male speakers.
$F2M$ and $M2F$ stand for cross-gender conversion. $F2M$ means that the source speaker is female, and the target speaker is male.
$M2F$ is the opposite of $F2M$.

\begin{figure*}[t]
\begin{center}
    \includegraphics[scale=0.60]{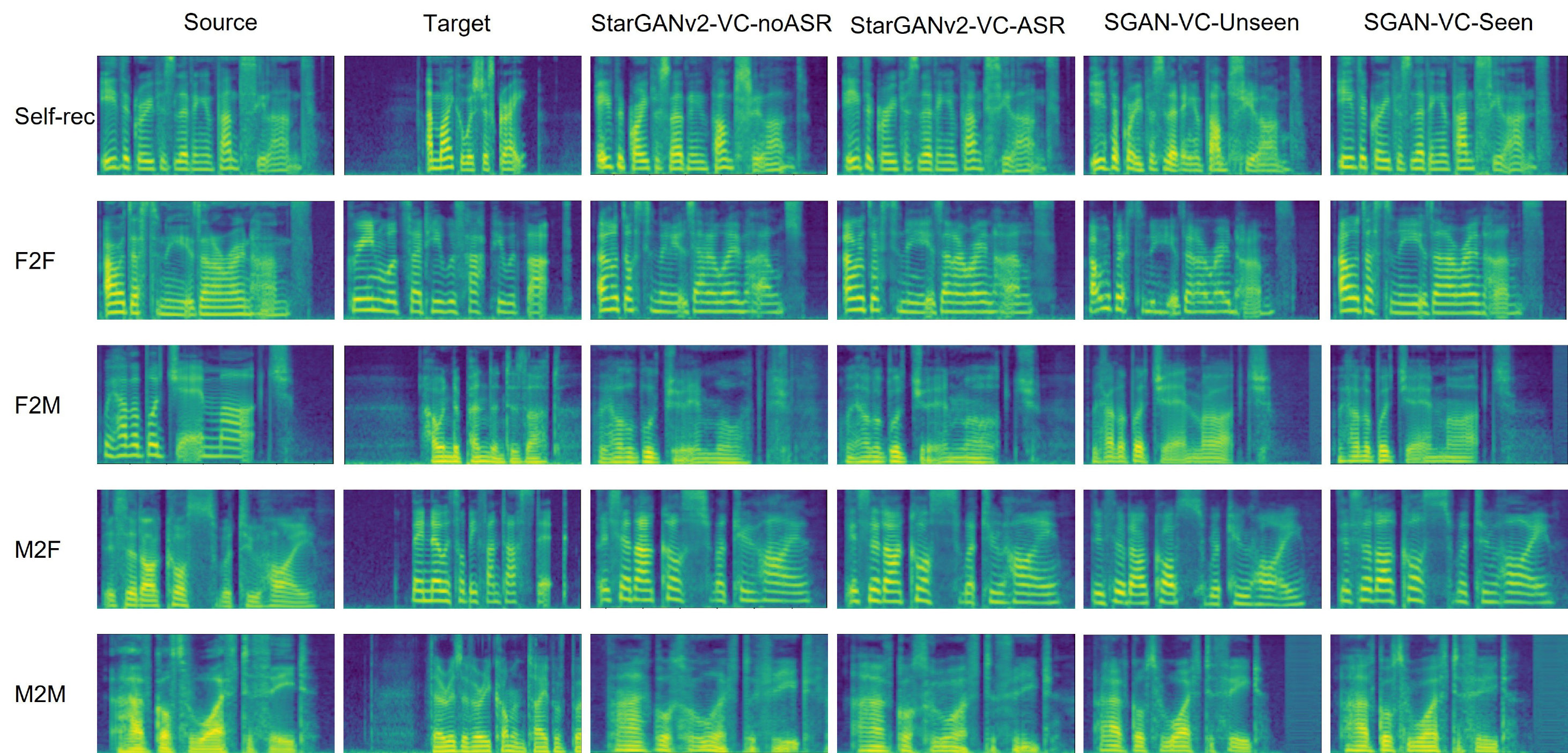}
\end{center}
\vspace{-.1in}
     \caption{Visualization of speeches transformed by different models. From top to bottom are self-reconstruction and four types of conversion. \textit{Self-rec} stands for self-reconstruction. From left to right are the source spectrogram, the target spectrogram, and the conversion results of the four methods StarGANv2-VC-noASR, StarGANv2-VC-ASR, SGAN-VC-Unseen, and SGAN-VC-Seen.
}
     \label{fig:vis}
\end{figure*}

For the seen experiment, the results of all methods are that the CLS of $F2M$ is higher than that of $M2F$.
As can be seen from Table~\ref{table:gender_cls}, our SGAN-VC-Seen outperforms StarGANv2-VC-ASR on all conversion types. 
Especially for $M2F$, our proposed method outperforms StarGANv2-VC-ASR by about $5\%$.
The above results validate our motivation for SGAN-VC: different subbands should be converted separately.
Simultaneously, our SGAN-VC can obtain comparable intelligibility results to StarGANv2-VC-ASR.
It shows that SGAN-VC has a strong capability to retain the source speech content while achieving high-similarity voice conversion.

\subsection{Visualization}
\label{sec:vis}
Figure~\ref{fig:vis} shows the visualization of the converted mel-spectrogram in the test set of VCTK Corpus.
StarGANv2-VC-ASR and StarGANv2-VC-noASR respectively represent the StarGANv2-VC model with or without ASR network assistance.
SGAN-VC-Seen and SGAN-VC-Unseen illustrate whether the speakers in the test set have appeared in the training set, respectively.
The first and second columns show the spectrogram samples of the source speaker and the target speaker.
Columns $3, 4, 5$, and $6$ display the converted mel-spectrograms for the four methods.
The first row shows the result of self-reconstruction, \ie, the same speaker is reconstructed according to its content feature and style feature.
Rows $2, 3, 4$, and $5$ depict four conversion types.
From the spectrograms of different conversion types, the conversion result of SGAN-VC has a high style similarity with the target speaker. 
Once the ASR network is removed, the source content of StarGANv2-VC is lost a lot, which can be clearly seen in the third line.
In terms of the voice content retention of the source speaker, SGAN-VC achieves a similar effect to StarGANv2-VC-ASR.

\setlength{\tabcolsep}{7pt}
\begin{table}
\small
\begin{center}
\begin{tabular}{lcccc}
\shline
Method & pMOS$\uparrow$ &  CLS$\uparrow$ &  CER$\downarrow$ & $mF0_{diff}\downarrow$ \\
\shline
 Ground truth & 3.484 & 96.60 \% & \multicolumn{1}{r}{5.27\%}  &  - \\
\cdashline{1-5}
 2s & 3.308 & 91.80 \% & 67.70 \% &  7.52 \\
 2.6s & \textbf{3.479} & \textbf{97.60} \% & \textbf{20.78} \% & \textbf{1.88}  \\
 3s & 3.390 & 97.30 \% & 79.61 \% &  7.47 \\
\hline
\end{tabular}
\end{center}
\vspace{-.1in}
\caption{The effect of audio clip duration on the VCTK20 dataset. All metrics are evaluated on 1000 conversion samples of random source and target pairs.}
\label{table:time effect vctk}
\end{table}

\subsection{Ablation Studies}

\noindent\textbf{Effect of the number of subbands.}
The number $n$ represents how many subbands the mel-spectrogram is divided.
When $n=1$, SGAN-VC takes the whole mel-spectrogram for voice conversion.
Due to the spatial characteristic of the mel-spectrogram, the implementations of several models differ slightly.
When the mel-spectrogram is not divided into subbands, since SGAN-VC converts holistic style information, there is no local information to disturb the converted style.
The overall style of the converted mel-spectrogram is close to the target speaker.
Therefore, in Table~\ref{table:num of part effects}, we can see that when $num=1$, CLS exceeds the results of other settings except for $num=4$. 
But in other evaluation metrics such as pMOS, the performance when $num=1$ is relatively low. Since only the overall style is converted, the phonemes of some words can not be converted naturally, resulting in a decrease in intelligibility. The results of CER and $mF0_{diff}$ are unsatisfying. When num is set to 2, since the spectrogram is only divided into two subbands, the local information is not divided reasonably enough, and the results of the three metrics are the lowest.
As $n$ increases, both CER and $mF0_{diff}$ improve because more detailed information is modeled. 
Intuitively, a larger $n$ allows the model to capture more detailed features.
The richer details make the converted style more realistic and accurate.
When $n=5$, all $4$ metrics dropped slightly.
Therefore, we use $n=4$ as the default choice for our framework, which balances the mining of contextual information with the appropriate size of the receptive field.

\noindent\textbf{Does the pitch-shift module work?}
As shown in the bottom two rows of Table \ref{table:object vctk}, in the experiment on the seen data, the performance of all four metrics is improved after adding the pitch-shift module. Similarly, in rows $4$ and $5$ of Table~\ref{table:object vctk}, in the experiment of unseen data, after adding the pitch-shift module, the three indicators of pMOS, CLS, and CER are improved. 
This indicates that the pitch-shift module fine-tunes the pitch during the generation of the converted speech. Of course, we can also see that the $mF0_{diff}$ metric does not improve in the unseen data experiments.
Because in the unseen data, the speakers in the test set are completely invisible during training. The parameters learned by the pitch-shift module have some slight gaps with the test data, resulting in a slight drop in the results. 
But we can also see from Table~\ref{table:vctk pitch difference} and Table~\ref{table:AISHELL3-84 pitch difference}, the pitch-shift module also has a certain improvement in the conversion of some speakers, such as, speakers $4$ and $7$ in Table~\ref{table:vctk pitch difference} and speakers $3, 4, 8$, and $10$ in Table~\ref{table:AISHELL3-84 pitch difference} in the results of unseen data experiments.

\noindent\textbf{Effect of the time length.} 
We finally test the effect of different time lengths respectively, as shown in \textbf{Table~\ref{table:time effect vctk}}. 
When the sound clip is too short or too long, the performance of the converted sound decreases. 
For short sounds, the model cannot convert the corresponding reference sounds, partially due to the lack of sufficient language and style information. 
Sound clips getting too long imply the model is getting too much content.
The fusion of style information and content information is under-thorough, which consequently limits the performance of the model.

\section{Conclusion}
\label{sec:conclusion}

We propose a subband-based generative adversarial network for non-parallel many-to-many voice conversion named SGAN-VC, which achieves state-of-the-art performance in terms of naturalness, content intelligibility, and style similarity.
SGAN-VC separately transfers the content in each frequency band from source style to target style.
Since the subbands are generated independently, the generative model can facilitate the timbre differences between different speakers.
Furthermore, the pitch-shift module enhances the interpretability of SGAN-VC.
Extensive experiments demonstrate the robustness and generalization of SGAN-VC.
The CER on the Mandarin and English datasets reaches or even exceeds the ASR network-assisted StarGANv2-VC. 
Especially in qualitative testing, for style similarity, SGAN-VC trained on unseen data is close to StarGANv2-VC on seen data.
Moreover, using HIFI-GAN as a vocoder makes our model suitable for real-time voice conversion applications.
SGAN-VC can realize the conversion of the arbitrary speaker without text transcription and parallel data, which is applicable in a wide range of scenarios.
In the future, we will focus on improving unseen speaker style similarity with limited training data.

\begin{figure}[t]
\begin{center}
    \includegraphics[width=1\linewidth]{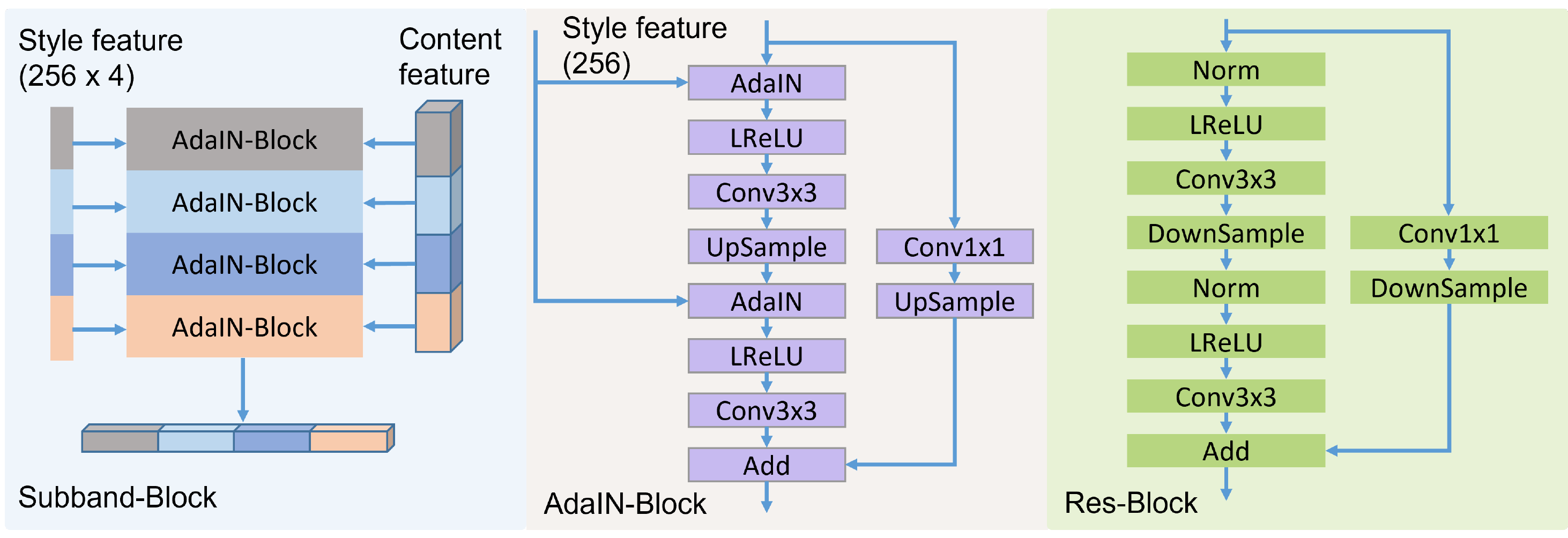}
\end{center}
\vspace{-.1in}
     \caption{Schematic diagram of network components. From left to right are Subband-Block, AdaIN-Block, and Res-Block. The norm in the content encoder and style encoder are Instance Normalization and BatchNorm, respectively.}
     \label{fig:blocks}
\end{figure}

\begin{figure}[t]
\begin{center}
    \includegraphics[width=1\linewidth]{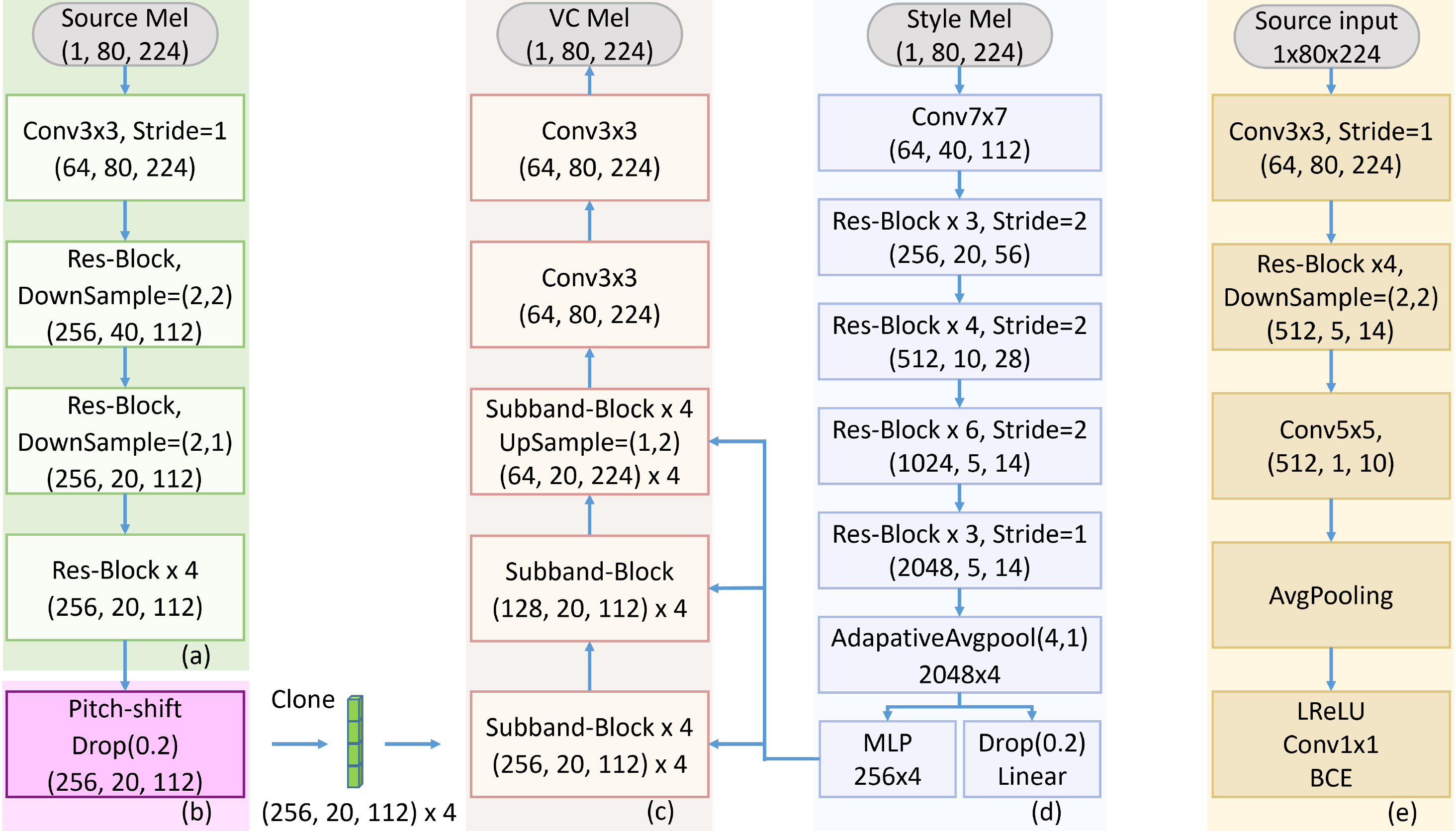}
\end{center}
\vspace{-.1in}
     \caption{A structure diagram of SGAN-VC. (a) Content Encoder, (b) Pitch-shift Module, (c) Decoder, (d) Style Encoder, (e) Discriminator. The Subband-Block is embedded in the Decoder.}
     \label{fig:struct}
\end{figure}

\appendices 
\section{NETWORK ARCHITECTURES}
\label{sec:appendices1}

SGAN-VC consists of a discriminator $Dis$ and a generative network, where the generator contains two encoders, \ie, content encoder $E_c$ and style encoder $E_s$, and a decoder $G$.
Refer to the practice in \cite{zheng2019joint,DBLP:conf/interspeech/LiZM21,huang2017arbitrary}, we mainly employ convolutions, Subband-Blocks, Res-Blocks, and AdaIN-Blocks to implement SGAN-VC, as shown in Figure~\ref{fig:blocks}.

\noindent\textbf{Content Encoder.} 
We adopt $E_c$ to extract the linguistic information of the source speaker.
Figure~\ref{fig:struct} (a) depicts the structure of $E_c$.
We first employ a $3\times3$ convolution to transform the input spectrogram into $64$ channels.
Then, we utilize $6$ Res-Blocks to extract content features.
In $E_c$, we employ Instance Normalization \cite{DBLP:journals/corr/UlyanovVL16} as our norm layers.
Since we only downsample twice vertically and once horizontally.
The size of the content feature is $(256, 20, 112)$.

\noindent\textbf{Pitch-shift module.} The pitch shift module is connected after the $E_c$ to fine-tune the pitch of the source speaker (Figure~\ref{fig:struct} (b)). It consists of five $5\times5$ convolutions, five Instance Normalization~\cite{DBLP:journals/corr/UlyanovVL16}, five LReLUs \cite{DBLP:journals/corr/XuWCL15}, one $1\times1$ convolution, and one Tanh activation function. 
Tanh is employed to normalize the output of the pitch-shift module to the $(-1,1)$ interval.
The output size of the pitch-shift module is a one-dimensional vector of dimension $112$, representing the offset for each time interval.
We utilize these offsets obtained from the pitch-shift module to fine-tune the pitch of the source speaker. 
Following the pitch-shift module \cite{DBLP:conf/interspeech/LiZM21}, we add a Dropout \cite{srivastava2014dropout} module to expand the generalization of SGAN-VC.
The final output of $E_c$ is a feature map of size $(256, 20, 112)$.

\noindent\textbf{Style Encoder.} 
\noindent{Figure~\ref{fig:struct} (d)} shows the architecture of the Style Encoder $E_s$. 
Following \cite{zheng2019joint}, we also adopt ResNet50 \cite{he2016deep} as the backbone of $E_s$.
Since the mel-spectrogram contains only one channel, we change the input dimension of the first convolutional layer of ResNet50 to $1$.
Finally, we obtain a feature map of size $(2048, 4, 14)$.
We utilize Adaptive Average Pooling to split the feature map into $4$ parts.
Each part represents the local style features of a subband.
Moreover, we perform average pooling on the feature map as a global feature representing the overall style.
To ensure overall style consistency, we concatenate each local feature with the global feature.
Ultimately, we get a $4\times4096$ feature vector. 
Each vector represents style information for a subband.
These feature vectors are fed into dropout and linear layers for style classification.
Simultaneously, after the integration and dimension reduction of the MLP module, the feature vectors are employed as the input of the Decoder. 
The MLP module consists of $3$ linear layers, $3$  Instance Normalization layers, and $3$ ReLU activation functions.
The dimension of the $4$ feature vectors is $256$.

\noindent\textbf{Decoder.}
The decoder is mainly composed of six Subband-Blocks and two $3\times3$ convolutions (Figure~\ref{fig:struct} (c)).
As can be seen in Figure 6, a Subband-Block contains 4 AdaIN-Blocks.
Each AdaIN-Block completes style transfer, according to the content feature from the source speaker and the style feature from the target speaker.
Each AdaIN-Block focuses on the conversion of the corresponding frequency band content.
To make the generated mel-spectrogram and the source spectrogram have the same size, AdaIN-Block upsamples once in the horizontal direction.
The last Subband-Block generates four feature maps of size $(64,20,224)$. We concatenate the feature maps together to form features of $(64,80,224)$.
Eventually, we utilize two $3\times3$ convolutions for feature fusion and generate converted mel-spectrograms.

\noindent\textbf{Discriminator.} To obtain the real/fake prediction (Figure~\ref{fig:struct} (e)), we deploy one $3\times3$ convolution, four Res-Blocks,  one $5\times5$ convolution, and one $1\times1$ convolution. 
BCE stands for Binary Cross-Entropy.

\ifCLASSOPTIONcaptionsoff
  \newpage
\fi


\bibliographystyle{IEEEtran}
\bibliography{IEEEabrv,mybib}

\vspace{-.4in}
\end{document}